\documentstyle[preprint,aps]{revtex}

\tightenlines

\begin{document}

\draft

\title{
A Detailed Determination of the A Priori Mixing Angles in\\
Non-Leptonic Decays of Hyperons
}

\author{
A.~Garc\'{\i}a
}
\address{
Departamento de F\'{\i}sica.
Centro de Investigaci\'on y de Estudios Avanzados del IPN\\
A.P. 14-740, M\'exico, D.F., 07000, MEXICO \\ 
}
\author{ 
R.~Huerta
}
\address{
Departamento de F\'{\i}sica Aplicada\\
Centro de Investigaci\'on y de Estudios Avanzados del IPN, Unidad M\'erida\\ 
A.P. 73, Cordemex. M\'erida, Yucat\'an 97310. MEXICO\\ 
}
\author{ 
G.~S\'anchez-Col\'on\footnote{
Permanent Address:
Departamento de F\'{\i}sica Aplicada.
Centro de Investigaci\'on y de Estudios Avanzados del IPN, Unidad M\'erida.
A.P. 73, Cordemex. M\'erida, Yucat\'an 97310. MEXICO.
}
}
\address{
Physics Department.
University of California\\
Riverside, California 92521-0413. U.S.A.\\
}

\date{May 11, 1998}

\maketitle

\begin{abstract}
Non-leptonic Decays of Hyperons can provide a detailed determination of the a
priori mixing angles that appear in physical hadrons in the approach in which
non-perturbative flavor and parity violations are present in tiny pieces of the
hadron mass operator.
The determination of such angles in these decays will provide a bench mark to test their necessary universality-like property in other types of decays.
Our main result is that the magnitudes of the a priori mixing angles can be determined quite accurately.
\end{abstract}

\pacs{
PACS Numbers:
13.40.Hg, 11.30.Er, 11.30.Hr, 12.60.-i
}

\section{Introduction}
\label{introduction}

Non-leptonic decays of hyperons (NLDH) can be understood by a
mechanism~\cite{LE-6129} due to non-perturbative flavor and parity mixings in
the physical hadrons and the intervention of the strong-interaction Yukawa
hamiltonian.
In Ref.~\cite{LE-6129} we have shown that this mechanism leads to the
predictions of the $|\Delta I|=1/2$ rule~\cite{marshak} for these decays, as
well as to numerical values of the so-called parity-conserving $B$ amplitudes
which are in good agreement with experiment.
If the new Yukawa coupling constants (YCC), that appear in the so-called
parity-violation $A$ amplitudes~\footnote{
We remind the reader that in the approach of a priori mixings in hadrons both
$A$ and $B$ amplitudes are actually parity and flavor conserving.
}
are assumed to have the same magnitudes as their ordinary counterparts, that appear in the $B$'s, then predictions for the $A$'s are obtained that also agree well with experiment.

All the YCC are constrained by their experimental values, the only free
parameters are the flavor and parity mixing angles $\sigma$, $\delta$, and
$\delta'$ that appear in physical hadrons.
We have refered to these angles as a priori mixings angles~\cite{LE-6129}, in
order to distinguish them from the perturbative ones that must be originated by
the intervention of the $W^\pm_\mu$ bosons.

In this paper we shall perform a detailed quantitative analysis, which was only
sketched in Ref.~\cite{LE-6129}.
Our main purpose will be not only to reproduce the experimental values of the
$A$'s and $B$'s but to establish as reliably as possible the values of $\delta$,
$\delta'$, and $\sigma$ in NLDH.
In Ref.~\cite{LE-6129} there was litle space for this latter task.
Such values are crucial to be able to proceed with the research program
discussed in the last section of Ref.~\cite{LE-6129}, namely, once their
values are determined in some type of decays they will be useful to test
their expected universality-like property in another type of decays.

In Sec.~\ref{formulas} we shall reproduce the expressions predicted for the
$A$ and $B$ amplitudes by the a priori mixings in hadrons approach.
In Sec.~\ref{data} the available experimental evidence that will be used in our
analysis will be discussed.
In Sec.~\ref{b} we shall study the $B$ amplitudes and the determination of
the a priori mixing angle $\sigma$ that accompanies them.
The $A$ amplitudes and their angles $\delta$ and $\delta'$ will be discussed in
Sec.~\ref{a}.
The simultaneous determination of the three angles will be considered in
Sec.~\ref{ab}.
The violation of the $|\Delta I|=1/2$ rule due to the breaking of isospin
invariance and its implications upon the values of $\sigma$, $\delta$, and
$\delta'$ will be studied in Sec.~\ref{sb}.
The last section is reserved for discussions and conclusions.

\section{A Priori Mixing Expressions for the $A$ and $B$ amplitudes}
\label{formulas}

For the sake of completeness and to introduce our notation, we shall reproduce
the expressions for the so-called parity-violating and parity-conserving $A$
and $B$ amplitudes, respectively, obtained if a priori flavor
and parity mixings exist in physical hadrons and the transition operator is the
strong-interaction Yukawa hamiltonian $H_Y$, namely,

\[
A_1
=
\delta'
\sqrt{\frac{3}{2}} g^{{}^{p,sp}}_{{}_{n,p\pi^-}} +
\delta
(
g^{{}^{s,ss}}_{{}_{\Lambda,pK^-}} - g^{{}^{s,pp}}_{{}_{\Lambda,\Sigma^+\pi^-}}
)
,
\]

\[
A_2
=
-\frac{1}{\sqrt{2}}
[
-\delta'
\sqrt{3}g^{{}^{p,sp}}_{{}_{n,n\pi^0}} +
\delta
(
g^{{}^{s,ss}}_{{}_{\Lambda,n\bar{K}^0}} -
\sqrt{3} g^{{}^{s,pp}}_{{}_{\Lambda,\Lambda\pi^0}} -
g^{{}^{s,pp}}_{{}_{\Lambda,\Sigma^0\pi^0}}
)
]
,
\]

\[
A_3
=
\delta
(
g^{{}^{s,ss}}_{{}_{\Sigma^-,nK^-}} +
\sqrt{\frac{3}{2}} g^{{}^{s,pp}}_{{}_{\Sigma^-,\Lambda\pi^-}} +
\frac{1}{\sqrt{2}} g^{{}^{s,pp}}_{{}_{\Sigma^-,\Sigma^0\pi^-}}
)
,
\]

\begin{equation}
A_4
=
-\delta'
g^{{}^{p,sp}}_{{}_{p,n\pi^+}} +
\delta
(
\sqrt{\frac{3}{2}} g^{{}^{s,pp}}_{{}_{\Sigma^+,\Lambda\pi^+}} +
\frac{1}{\sqrt{2}} g^{{}^{s,pp}}_{{}_{\Sigma^+,\Sigma^0\pi^+}}
)
,
\label{aes}
\end{equation}

\[
A_5
=
- \delta'
g^{{}^{p,sp}}_{{}_{p,p\pi^0}} -
\delta
(
\frac{1}{\sqrt{2}} g^{{}^{s,ss}}_{{}_{\Sigma^+,p\bar{K}^0}} +
g^{{}^{s,pp}}_{{}_{\Sigma^+,\Sigma^+\pi^0}}
)
,
\]

\[
A_6
=
\delta'
g^{{}^{p,sp}}_{{}_{\Sigma^-,\Lambda\pi^-}} +
\delta
(
g^{{}^{s,ss}}_{{}_{\Xi^-,\Lambda K^-}} + 
\sqrt{\frac{3}{2}} g^{{}^{s,pp}}_{{}_{\Xi^-,\Xi^0\pi^-}}
)
,
\]

\[
A_7
=
\frac{1}{\sqrt{2}}
[
\delta'
(
\sqrt{3} g^{{}^{p,sp}}_{{}_{\Lambda,\Lambda\pi^0}} +
g^{{}^{p,sp}}_{{}_{\Sigma^0,\Lambda\pi^0}}
)
+
\delta
(
- g^{{}^{s,ss}}_{{}_{\Xi^0,\Lambda\bar{K}^0}} + 
\sqrt{3} g^{{}^{s,pp}}_{{}_{\Xi^0,\Xi^0\pi^0}}
)
]
,
\]

\noindent
and

\[
B_1
=
\sigma
(
- \sqrt{\frac{3}{2}} g_{{}_{n,p\pi^-}} +
g_{{}_{\Lambda,pK^-}} - g_{{}_{\Lambda,\Sigma^+\pi^-}}
)
,
\]

\[
B_2
=
-\frac{1}{\sqrt{2}}
\sigma
(
\sqrt{3}g_{{}_{n,n\pi^0}} +
g_{{}_{\Lambda,n\bar{K}^0}} -
\sqrt{3} g_{{}_{\Lambda,\Lambda\pi^0}} -
g_{{}_{\Lambda,\Sigma^0\pi^0}}
)
,
\]

\[
B_3
=
\sigma
(
g_{{}_{\Sigma^-,nK^-}} +
\sqrt{\frac{3}{2}} g_{{}_{\Sigma^-,\Lambda\pi^-}} +
\frac{1}{\sqrt{2}} g_{{}_{\Sigma^-,\Sigma^0\pi^-}}
)
,
\]

\begin{equation}
B_4
=
\sigma
(
g_{{}_{p,n\pi^+}} +
\sqrt{\frac{3}{2}} g_{{}_{\Sigma^+,\Lambda\pi^+}} +
\frac{1}{\sqrt{2}} g_{{}_{\Sigma^+,\Sigma^0\pi^+}}
)
,
\label{bes}
\end{equation}

\[
B_5
=
\sigma
(
g_{{}_{p,p\pi^0}} -
\frac{1}{\sqrt{2}} g_{{}_{\Sigma^+,p\bar{K}^0}} -
g_{{}_{\Sigma^+,\Sigma^+\pi^0}}
)
,
\]

\[
B_6
=
\sigma
(
- g_{{}_{\Sigma^-,\Lambda\pi^-}} +
g_{{}_{\Xi^-,\Lambda K^-}} + 
\sqrt{\frac{3}{2}} g_{{}_{\Xi^-,\Xi^0\pi^-}}
)
,
\]

\[
B_7
=
\frac{1}{\sqrt{2}}
\sigma
(
- \sqrt{3} g_{{}_{\Lambda,\Lambda\pi^0}} -
g_{{}_{\Sigma^0,\Lambda\pi^0}} -
g_{{}_{\Xi^0,\Lambda\bar{K}^0}} + 
\sqrt{3} g_{{}_{\Xi^0,\Xi^0\pi^0}}
)
.
\]

\noindent
The subindeces $1,\dots,7$ correspond to
$\Lambda\rightarrow p\pi^-$,
$\Lambda\rightarrow n\pi^0$, 
$\Sigma^-\rightarrow n\pi^-$, 
$\Sigma^+\rightarrow n\pi^+$, 
$\Sigma^+\rightarrow p\pi^0$, 
$\Xi^-\rightarrow \Lambda\pi^-$,
and 
$\Xi^0\rightarrow \Lambda\pi^0$,
respectively.
The YCC in the $B$'s are the ordinary ones, while the YCC in the $A$'s are new
ones.
In the latter, the upper indeces serve as a reminder of the parities of the
parity eigenstates involved.
$\delta$, $\delta'$, and $\sigma$ are the mixing angles that appear in the a
priori mixed hadrons.
We remind the reader that these physical hadrons are obtained, given our
current inability to compute well with QCD, following the ansatz discussed in
Ref.~\cite{LE-6129}.
The a priori mixed hadrons thus obtained and used to obtain Eqs.~(\ref{aes})
and (\ref{bes}) are

\[
K^+_{ph} =
K^+_{0p} - \sigma \pi^+_{0p} - \delta' \pi^+_{0s}
+ \cdots
,
\] 

\[
K^0_{ph} = 
K^0_{0p} +
\frac{1}{\sqrt{2}} \sigma \pi^0_{0p} + \frac{1}{\sqrt{2}} \delta' \pi^0_{0s}
+ \cdots
,
\]
 
\begin{equation}
\pi^+_{ph} = 
\pi^+_{0p} + \sigma K^+_{0p} - \delta K^+_{0s}
+ \cdots
,
\label{mph}
\end{equation}
 
\[
\pi^0_{ph} =
\pi^0_{0p} -
\frac{1}{\sqrt{2}} \sigma ( K^0_{0p} + \bar{K}^0_{0p} ) +
\frac{1}{\sqrt{2}} \delta ( K^0_{0s} - \bar{K}^0_{0s} )
+ \cdots
,
\]
 
\[
\pi^-_{ph} =
\pi^-_{0p} + \sigma K^-_{0p} + \delta K^-_{0s}
+ \cdots
,
\]
 
\[
\bar{K}^0_{ph} =
\bar{K}^0_{0p} + \frac{1}{\sqrt{2}} \sigma \pi^0_{0p} -
\frac{1}{\sqrt{2}} \delta'\pi^0_{0s}
+ \cdots
,
\]

\[
K^-_{ph} =
K^-_{0p} - \sigma \pi^-_{0p} + \delta' \pi^-_{0s}
+ \cdots
,
\]

\noindent
and,
 
\[
p_{ph} = 
p_{0s} - \sigma \Sigma^+_{0s} - \delta \Sigma^+_{0p}
+ \cdots
,
\]
 
\[
n_{ph} = 
n_{0s} +
\sigma ( \frac{1}{\sqrt{2}} \Sigma^0_{0s} + \sqrt{\frac{3}{2}} \Lambda_{0s}) +
\delta ( \frac{1}{\sqrt{2}} \Sigma^0_{0p} + \sqrt{\frac{3}{2}} \Lambda_{0p} )
+ \cdots
,
\]
             
\[
\Sigma^+_{ph} =
\Sigma^+_{0s} + \sigma p_{0s} - \delta' p_{0p}
+ \cdots
,
\]
 
\begin{equation}
\Sigma^0_{ph} =
\Sigma^0_{0s} +
\frac{1}{\sqrt{2}} \sigma ( \Xi^0_{0s}- n_{0s} ) +
\frac{1}{\sqrt{2}} \delta \Xi^0_{0p} + \frac{1}{\sqrt{2}} \delta' n_{0p}
+ \cdots
,
\label{bph}
\end{equation}
 
\[
\Sigma^-_{ph} = \Sigma^-_{0s} + \sigma \Xi^-_{0s} + \delta \Xi^-_{0p} 
+ \cdots
,
\]

\[
\Lambda_{ph} = 
\Lambda_{0s} + 
\sqrt{\frac{3}{2}} \sigma ( \Xi^0_{0s}- n_{0s} ) +
\sqrt{\frac{3}{2}} \delta \Xi^0_{0p} + 
\sqrt{\frac{3}{2}} \delta' n_{0p}
+ \cdots
,
\]
 
\[
\Xi^0_{ph} =
\Xi^0_{0s} -
\sigma
( \frac{1}{\sqrt{2}} \Sigma^0_{0s} + \sqrt{\frac{3}{2}} \Lambda_{0s} ) +
\delta'
( \frac{1}{\sqrt{2}} \Sigma^0_{0p} + \sqrt{\frac{3}{2}} \Lambda_{0p} )
+ \cdots
,
\]

\[
\Xi^-_{ph} =
\Xi^-_{0s} - \sigma \Sigma^-_{0s} + \delta' \Sigma^-_{0p}
+ \cdots
.
\]

\noindent
The dots stand for other mixings not used in obtaining Eqs.~(\ref{aes}) and
(\ref{bes}).
The subindices naught, $s$, and $p$ mean flavor, positive, and negative parity
eigenstates, respectively.

The main qualitative and already semi-quantitative result obtained with the
above expressions for the $A$'s and $B$'s are the predictions of the
$|\Delta I|=1/2$ rule for NLDH, when $H_Y$ is assumed to be an isospin $SU(2)$
invariant-operator.
In this symmetry limit, as discussed in detail in Ref.~\cite{LE-6129}, one
obtains the equalities~\cite{dumbrajs}

\begin{equation}
A_2 = - \frac{1}{\sqrt{2}} A_1, \ \ \ \ \ \ 
A_5 = \frac{1}{\sqrt{2}} ( A_4 - A_3 ), \ \ \ \ \ \ 
A_7 = \frac{1}{\sqrt{2}} A_6,
\label{sla}
\end{equation}

\noindent
and

\begin{equation}
B_2 = - \frac{1}{\sqrt{2}} B_1, \ \ \ \ \ \ 
B_5 = \frac{1}{\sqrt{2}} ( B_4 - B_3 ), \ \ \ \ \ \ 
B_7 = \frac{1}{\sqrt{2}} B_6.
\label{slb}
\end{equation}

The $SU(2)$ symmetry limit of the YCC leads to the equalities
$g_{{}_{p,p\pi^0}}=-g_{{}_{n,n\pi^0}}=g_{{}_{p,n\pi^+}}/{\sqrt 2}
=g_{{}_{n,p\pi^-}}/{\sqrt 2}$,
$g_{{}_{\Sigma^+,\Lambda\pi^+}}=g_{{}_{\Sigma^0,\Lambda\pi^0}}
=g_{{}_{\Sigma^-,\Lambda\pi^-}}$,
$g_{{}_{\Lambda,\Sigma^+\pi^-}}=g_{{}_{\Lambda,\Sigma^0\pi^0}}$,
$g_{{}_{\Sigma^+,\Sigma^+\pi^0}}=-g_{{}_{\Sigma^+,\Sigma^0\pi^+}}
=g_{{}_{\Sigma^-,\Sigma^0\pi^-}}$,
$g_{{}_{\Sigma^0,pK^-}}=g_{{}_{\Sigma^-,nK^-}}/{\sqrt 2}
=g_{{}_{\Sigma^+,p\bar K^0}}/{\sqrt 2}$,
$g_{{}_{\Lambda,pK^-}}=g_{{}_{\Lambda,n\bar K^0}}$,
$g_{{}_{\Xi^0,\Xi^0\pi^0}}=g_{{}_{\Xi^-,\Xi^0\pi^-}}/{\sqrt 2}$,
$g_{{}_{\Xi^-,\Lambda K^-}}=-g_{{}_{\Xi^0,\Lambda \bar K^0}}$,
and
$g_{{}_{\Lambda,\Lambda \pi^0}}=0$.
It is these equalities that lead to Eqs.~(\ref{slb}) when they are used in
Eqs.~(\ref{bes}).
Similar relations are valid within each set of upper indeces, e.\ g.,
$g^{{}^{p,sp}}_{{}_{p,p\pi^0}}=-g^{{}^{p,sp}}_{{}_{n,n\pi^0}}$, etc.\ when
$SU(2)$ symmetry is applied to the new YCC.
The equalities thus obtained lead to Eqs.~(\ref{sla}) when they are used in
Eqs.~(\ref{aes}).

In the next section we shall discuss the available experimental evidence on
NLDH, on the YCC, and the relevance of the signs of the $A$ and $B$ amplitudes.

\section{Experimental data and the signs of the $A$ and $B$ amplitudes}
\label{data}

The experimental data~\cite{pdg} on the seven NLDH we are concerned with here
come in the form of decay rates $\Gamma_i$ and spin asymmetries $\alpha_i$ and
$\gamma_i$ ($i=1,\dots,7$)~\footnote{
We find it convenient to use the $\gamma_i$-asymmetries, instead of the angle
$\phi_i$.
The experimental correlation pointed out in Ref.~\cite{pdg} has a
negligible effect in our analysis.
}.
These data are listed in Table~\ref{tablai}.

Absorbing certain kinematical and overall factors, these observables take the
particularly simple forms $\Gamma_i=S^2_i+P^2_i$,
$\alpha_i=2S_iP_i/(S^2_i+P^2_i)$, and $\gamma_i=(S^2_i-P^2_i)/(S^2_i+P^2_i)$.
We shall ignore final state interactions and assume $CP$-invariance; thus, each
amplitude is real.
$S_i$ are proportional to the $A_i$ and $P_i$ are proportional to the $B_i$.
It is also customary to quote experimental values for all the amplitudes.
This we do too in Table~\ref{tablai}.
However, the determination of the signs of the amplitudes requires a detailed
discussion.

In a plane whose cartesian axes correspond to $S_i$ and $P_i$, $\Gamma_i$
represents a circunference and $\alpha_i$ a hyperbola.
There are four intersections between these two curves.
These four solutions are such that one is equal to another one up to an overall
sign; so there are actually only two solutions up to such overall signs.
In addition one of the two solutions becomes the other one by interchanging the
magnitudes of $S_i$ with $P_i$ (or of $A_i$ with $B_i$).
The role of $\gamma_i$ is to determine the relative magnitudes between $S_i$
and $P_i$ (or between $A_i$ and $B_i$).
Their relative sign is fixed by $\alpha_i$.
Therefore, the relative sign and the relative magnitudes between $A_i$ and
$B_i$ are unique, but their overall signs cannot be experimentally
determined.

We have freedom to chose the overall signs, but Eqs.~(\ref{sla}) and (\ref{slb})
impose many restrictions, which are very important because they are predictions
independent of the particular values of the YCC and the a priori mixing angles.
They are part of the predictions of the $|\Delta I|=1/2$ rule.
For definiteness we shall asign the overall signs to the $B_i$ amplitudes.

Since there are seven $B_i$ amplitudes and two signs, we have $2^7$
possibilities.
However, the relative signs between $B_1$ and $B_2$, $B_4$ and $B_5$, and $B_6$
and $B_7$ are fixed by Eqs.~(\ref{slb}) and the fact that $|B_4|$,
$|B_5|\gg|B_3|$.
They are required to be negative, positive, and positive, respectively.
Our choice is then limited to $2^4=16$ posibilities.
There is still another limitation.
Eqs.~(\ref{sla}) and the fact that $|A_3|$, $|A_5|\gg|A_4|$ require that the
relative sign between $A_3$ and $A_5$ be opposite.
In addition, from Table~\ref{tablai}, we see that $\alpha_3<0$ and thus that
the relative sign between $B_3$ and $A_3$ must be negative.
Therefore, $B_3$ and $A_5$ must have the same sign and since $B_5$ has the same
sign as $B_4$, the relative sign between $B_3$ and $B_4$ is fixed to be the
same as the relative sign between $B_5$ and $A_5$, i.\ e., as the sign of
$\alpha_5$.
Since $\alpha_5<0$, the sign between $B_3$ and $B_4$ must be negative.
Clearly, we are left with only $2^3=8$ possiblities, out of the initial $2^7$.
These eight possibilities we shall apply to $B_1$, $B_3$, and $B_6$.
So, for example, if we choose $B_1>0$, $B_3<0$, and $B_6<0$, then we have fixed
$B_2<0$, $B_4>0$, $B_5>0$, and $B_7<0$.
Then, from the above discussion we also have $A_3>0$, $A_5<0$.
Knowing that $\alpha_1>0$, $\alpha_4>0$, and $\alpha_6<0$ we are forced to take
$A_1>0$, $A_4>0$, and $A_6>0$.
Finally the signs of $A_2<0$ and $A_7>0$ are fixed by Eqs.~(\ref{sla}).
Proceeding this way we can form the remainig seven choices.
All the sign possibilities are collected in Table~\ref{tablaii}.

Notice that since the relative signs between $A_1$ and $A_2$ and $B_1$ and $B_2$
are fixed by Eqs.~(\ref{sla}) and (\ref{slb}) (both negative), once the
relative sign between $A_1$ and $B_1$ is fixed experimentally by $\alpha_1$,
the relative sign between $A_2$ and $B_2$ is also fixed and it is fixed to be
the same as the sign of $\alpha_1$.
That is, irrespective of the above freedom to choose overall signs,
Eqs.~(\ref{sla}) and (\ref{slb}) predict that the sign of $\alpha_2$ must be
the same of $\alpha_1$.
Analogous remarks apply to $A_6$, $A_7$, $B_6$, and $B_7$,
Eqs.~(\ref{sla}) and (\ref{slb}) predict that the sign of $\alpha_7$ must be
the same of $\alpha_6$.
These predictions are very general, they are independent of the particular
values of $\delta$, $\delta'$, and $\sigma$ and of the particular isospin
symmetry-limit values of the YCC that may appear in Eqs.~(\ref{aes}) and
(\ref{bes}).
In Table~\ref{tablai}, we can verify that these two predictions are indeed
experimentally confirmed.

To close this section let us list in Table~\ref{tablaiii}, for easy later
reference, the experimental values of the ordinary YCC currently available.
Only the squares of five couplings are quoted in Ref.~\cite{dumbrajs}, but we
shall need two more $g_{{}_{\Xi^0,\Xi^0\pi^0}}$ and
$g_{{}_{\Xi^-,\Lambda K^-}}$.
Also, their relative signs are important.
In as-much-as strong-flavor $SU_3$, broken as it is, is a reliable symmetry, we
shall assume that the relative signs are fixed by this symmetry.
Along these lines, we shall then assume that $g_{{}_{\Xi^0,\Xi^0\pi^0}}$ and
$g_{{}_{\Xi^-,\Lambda K^-}}$ can be estimated by their $SU_3$ relationship, but
assign to them an error bar allowing for variations of some $30\%$ around such
values used as central values.
The values entered into Table~\ref{tablaiii} are normalized to the pion-nucleon
YCC (assumed to be positive).

The data of Table~\ref{tablai} should expected to be very reliable, they have
been obtained through many experiments which have shown very acceptable
agreement with one another.
The experimental values of the YCC of Table~\ref{tablaiii} may not be so
stable.
They are model dependent to an extent which is difficult to assess and the
attempts to determine them have not always been free of controversy.
It should not be surprising that these data show in future determinations some
important changes.
However, it should be emphasized that they show reasonable consistency with
broken strong $SU_3$ symmetry at the expected 20--30$\%$ level.
It is probably this last remark that provides the best line of judgement in
their use.

\section{Determination of the $B$ amplitudes and the angle $\sigma$}
\label{b}

The predictions for the so-called parity conserving
amplitudes in the a priori mixed hadron approach, given in
Eqs.~(\ref{bes}), require that the several YCC that appear in them be
identified with the ordinary ones determined in strong-interaction physics.
They are not free parameters, they are constrained by the currently
experimental values displayed in Table~\ref{tablaiii}.
In contrast, the angle $\sigma$ remains unconstrained.
We do not have any theoretical argument which could help us fix it or even
loosely bound it.
We must leave it as a free parameter and extract its value from the comparison
of Eqs.~(\ref{bes}) with their counterparts in Table~\ref{tablai}.

As discussed in the last section, the phases of the $B$'s cannot be determined,
but out of all the possible choices only the eight ones displayed in
Table~\ref{tablaii} turned out to be acceptable.
We cannot tell in advance which of these eight choices can be reproduced by
Eqs.~(\ref{bes}), so we must try them all.
It turns out that four of them are the best reproduced.
We collect in Table~\ref{tablaiv} all the predictions of Eqs.~(\ref{bes}) in
these four choices, along with the values of the YCC and the angle $\sigma$.

Let us discuss the results obtained.
The experimental values of the $B$'s are very well reproduced.
The YCC come out quite reasonably close the values of Table~\ref{tablaiii}.
Most importantly, all the $SU_3$ signs are reproduced.
A very interesting feature is that the value of the only free parameter
$\sigma$ remains quite stable along the four choices of signs of the
experimental $B$'s.

The results obtained are good enough to conclude that a priori mixings in
hadrons, not only yield the $|\Delta I|=1/2$ rule predictions for the parity
conserving amplitudes of NLDH, but also provide a very good framework for their
detailed description in terms of only one free parameter.

\section{
Determination of the $A$ amplitudes and the angles $\delta$ and $\delta'$
}
\label{a}

The predictions of a priori mixings in hadrons for the so-called parity
violating amplitudes $A$ are given in Eqs.~(\ref{aes}).
New YCC are involved in them and this is indicated by the indeces $s$ and $p$
attached.
Although Eqs.~(\ref{aes}) provide a framework, we face a practical difficulty.
Due to our current inability to compute well with QCD, we are unable to obtain
the theoretical values of these new YCC and accordingly we must leave them as
free parameters in order to reproduce the experimental $A$'s of
Table~\ref{tablai}.
However, this is not good enough because there are more parameters than $A$
amplitudes.
If we try this latter way we simply learn nothing and we cannot determine the
angles $\delta$ and $\delta'$.
If we want to proceed, we must introduce constraints on these YCC by making
educated guesses.

Since QCD is assumed to be common to the positive and negative parity quarks of
the anzats we have used for guidance, one may expect that the new YCC are some
how related to the ordinary ones.
Specifically, one may reasonably expect that the magnitudes of the YCC are
the same as the magnitudes of the ordinary ones.
Their signs may differ however.
We shall impose these constraints on the new YCC.
Although, this way they are not free parameters anymore, we must still face many
possibilities since there are two signs to be chosen for each one of the new
YCC.
Therefore, we should perform a systematic analysis allowing for each possible
choice of relative signs between the new and the ordinary sets of YCC.
This analysis presents no essential difficulty although it is a tedious one.

The results of this analysis are very interesting.
Not all of the choices are allowed.
As a matter of fact, most of them are ruled out, but still many of the choices
remain possible, one out of every five.
That is, out of the initial 256 possibilities only about 50 remain.
We shall not display them all but we shall mention their most important
features.
In each one of these 50 or so possibilities the $A$'s are always reasonably well
reproduced (with the eight overall signs of Table~\ref{tablaii} taken into
consideration) and the magnitudes obtained for the new YCC come out also close
to the experimental magnitudes of Table~\ref{tablaiii}, but the most important
result is that the values of the $\delta$ and $\delta'$ angles show a
remarkable systematics: they always come out in either one of two groups.
They either take values around
$\delta = 0.10\times 10^{-6}$
and
$\delta' = 0.04\times 10^{-6}$
or around
$\delta = 0.15\times 10^{-6}$
and
$\delta' = 0.30\times 10^{-6}$.

The main conclusion of this analysis is, then, that (i) the $A$'s can be
reproduced in detail and (ii) the possible values of $\delta$ and $\delta'$ are
reduced to only two sets.

Even if it is not necessary to display all of the many cases of the above
analysis, it is convenient to show some of them.
This we do in Table~\ref{tablav}.
The cases displayed will be quite relevant in what folows.
The predictions for the $A$'s should be compared with the corresponding
experimental ones in Table~\ref{tablaii} and the YCC should be compared with
the values in Table~\ref{tablaiii}.
One can see in these comparisons that the results obtained are quite acceptable
and, therefore, that Eqs.~(\ref{aes}) can describe the $A$ amplitudes fairly
well when the new YCC are constrained in their magnitudes by the values of
Table~\ref{tablaiii}.
One can also appreciate that $\delta$ and $\delta'$ are determined within two
sets of values, in accordance with their systematic behavior mentioned before.

\section{
Predictions for the experimental observables and simultaneous determination of
$\delta$, $\delta'$, and $\sigma$
}
\label{ab}

Predicting the $A$ and $B$ amplitudes separately is really an intermediate
step, one must do more and proceed to predict the complete collection of
experimental observables of Table~\ref{tablai}.
This represents a substantially more stringent test of Eqs.~(\ref{aes}) and
(\ref{bes}) as we shall presently see.

We have seen in the last sections that there seems to be many solutions for
Eqs.~(\ref{aes}) and (\ref{bes}) to describe NLDH.
These many solutions arise in the choices for the overall signs of the $A$'s and
$B$'s and are increased by the free relative signs between the new and the
ordinary YCC.
In addition, although the new YCC were constrained in their magnitudes by the
experimental values of Table~\ref{tablaiii}, the error bars allow small
differences between the magnitudes of YCC in going from $A$ to $B$ amplitudes.
This can be observed by comparing the corresponding entries in
Tables~\ref{tablaiv} and \ref{tablav}.
Therefore, the strict equality of the magnitudes of the new and ordinary YCC
can only be enforced by reproducing the complete set of experimental
observables.
This should then reduce appreciably the number of solutions found in the
previous sections.

We must then perform both of the systematic searches of Secs.~\ref{b} and
\ref{a} simultaneously while using all the data for the observables $\Gamma_i$,
$\alpha_i$, and $\gamma_i$ of Table~\ref{tablai}.
After performing this analysis, a very important result is obtained: the best
description of experimental data is reduced to three cases.
Most of the possibilities for the $B$'s and $A$'s are ruled out and a few
remain which are not too bad but are no longer as good as they appeared at
first.
These three cases are displayed in Tables~\ref{tablavi} and \ref{tablavii}.

In these three cases the experimental data are very well reproduced and the
YCC are also very well reproduced in the first two cases (from left to right
in Table~\ref{tablavii}), while in the third case (to the right of
Table~\ref{tablavii}) one can observe some variations which are not
negligible.
This last observation, however, must be taken with care, since as we remarked
at the end of Sec.~\ref{data}, the experimental values of Table~\ref{tablaiii}
may change in the future because the experimental determination of the YCC is
quite difficult.
With this in mind, we find that the third case is acceptable.

It must be pointed out the stability of the three values obtained for
$\sigma$.
It must also be pointed out that $\delta$ and $\delta'$ still fall into either
one of the two sets found in Sec.~\ref{a}.
The a priori mixing angles are fairly well determined whether one uses the
experimental amplitudes or the experimental observables.                      

\section{
Violations of $|\Delta I|=1/2$ rule predictions through $SU(2)$ symmetry
breaking
}
\label{sb}

It is well known that experimentally the predictions of the $|\Delta I|=1/2$
rule are not exact.
In the case of a priori mixings in hadrons the violations of these predictions
will come by the breaking of the $SU(2)$ strong-flavor symmetry, which was
introduced by assuming that the Yukawa hamiltonian was an $SU(2)$ scalar.
It is, therefore, necessary to explore the effect of such breaking.

In this section we shall let the YCC that appear in Eqs.~(\ref{aes}) and
(\ref{bes}) to differ from their $SU(2)$ symmetry limit.
However, we shall allow for only small differences from this limit by
constraining such changes to remain at the few percent level.
Stronger variations will not be considered.
This analysis leads to Tables~\ref{tablaviii} and \ref{tablaix}.

In going through Tables~\ref{tablaviii} and \ref{tablaix} and after comparing
them with Tables~\ref{tablavi} and \ref{tablavii}, respectively, one can
observe that changes of a few percent from the $SU(2)$ symmetry limit of the
YCC allow the predictions of Eqs.~(\ref{aes}) and (\ref{bes}) to describe the
experimental data even better than before.
The YCC that go into these predictions come out very reasonable.
Also the variations observed in the third case (from left to right) in
Table~\ref{tablavii} are milder now.
The main effect of these small changes is seen in the error bars of the a priori
mixing angles, especially in the second and third cases.

We must conclude that the determination of the a priori mixing angles is
affected by $SU(2)$ breaking corrections and we should avoid overestimating the
values obtained for them in NLDH.
One must then be cautious and quote the values of $\sigma$, $\delta$, and
$\delta'$ affected by $SU(2)$ breaking, namely, those of Table~\ref{tablaix}.

\section{Discussions and conclusions}
\label{conclusions}

Throughout the last four sections we have performed a very detailed analysis of
the ability of the a priori mixings in hadrons to describe NLDH.
This description comes in four levels.

First, there are very general features which cover the predictions of the
$|\Delta I|=1/2$ rule.
They are independent of the values of the a priori mixing angles and of the
YCC and they will be violated only by $SU(2)$ symmetry breakings.
Since these breakings are very small, these predicitons are quite accurate, as
is experimentally the case.

Second, the so-called parity-conserving $B$ amplitudes must
be described using the values of the YCC observed in the strong-interactions of
hyperons and mesons.
Only a free parameter remains.
The results obtained for the $B$ amplitudes are very good, as was discussed in
Sec.~\ref{b}.
No new assumptions had to be introduced.

Third, in order to describe the so-called parity-violating $A$
amplitudes one has to introduce new assumptions because new
YCC are involved.
It is reasonable to expect that these YCC are not completely independent of the
ordinary YCC.
We have introduced an educated guess based on the original motivation that led
to the ansatz which we introduced to guide ourselves for the practical
implementation of a priori mixings in hadrons.
Since QCD is common to both the positive and negative parity quarks used in our
ansatz, one may expect that the magnitudes of the new and ordinary YCC be the
same.
This still leaves the relative signs open.
We performed a thorough study in Sec.~\ref{a}, showing that the $A$'s can be
well described.
The two free angles $\delta$ and $\delta'$ showed a remarkable stability and
always stayed close to either one of two sets of values.

Fourth, the very many possibilities allowed at the third level are very much
reduced when trying to cover simultaneously all of the available data on the
experimental observables in NLDH.
The best cases are reduced to three and were displayed in the tables of
Sec.~\ref{ab}.

Since it is well known that the predictions of the $|\Delta I|=1/2$ rule are
not exact experimentally and in order to complete our analysis, in
Sec.~\ref{sb} we allowed for the presence of small $SU(2)$ violations in the
YCC.
This exercise taught us that one should be somewhat cautious when
determinig the a priori mixing angles, but otherwise these
small violations are seen to improve even more the agreement between
Eqs.~(\ref{aes}) and (\ref{bes}) and experiment.

There is another point whose discussion we wanted to leave to the end of this
paper.
We did not commit ourselves in any way about the relative signs between
the new and the ordinary YCC.
In this respect, there are several interesting observations we wish to make.
Again, since QCD is assumed to be common to the positive and negative parity
quarks of our ansatz, one could expect that the mechanism that assigns
strong-flavors to positive-parity hyperons and negative-parity mesons be common
to negative-parity baryons and positive-parity mesons.
That is, it could be possible that the latter hadrons come in $SU(3)$ octets
too, albeit, different octets than those of the former ones.
If this were to be the case then one could expect that certain relative signs of
the new YCC be the same as the relative signs of the corresponding ordinary
YCC.
We can distinguish three groups of the new YCC according to the indeces
$(p,sp)$, $(s,ss)$, and $(s,pp)$ in Eqs.~(\ref{aes}).
One could expect that the relative signs of the new YCC within each one of
these groups be the same as the relative signs between the corresponding
ordinary YCC of Table~\ref{tablaiii}.
One can observe that this is indeed the case in the third solution of
Sec.~\ref{ab}.
This can be taken as an indication that it could make sense to go beyond the
assumption of only equating the magnitudes of the new and old YCC.

To close this paper let us make some comments about the values of the a priori
mixing angles.
Perhaps the most striking result of our detailed analysis is the relatiive
stability of the values obtained for them throughout all the cases considered.
The best values we have obtained for them are those of Table~\ref{tablaix}.
We have there three sets of values which even if they are not quite unique they
are, however, very close to one another.
In view of the last remarks, one might be inclined to prefer the set of the
second solution in this table, namely,

\begin{eqnarray}
\sigma & = & (4.9\pm 2.0)\times 10^{-6}
\nonumber \\
|\delta| & = & (0.22\pm 0.09)\times 10^{-6}
\label{ang} \\
|\delta'| & = & (0.26\pm 0.09)\times 10^{-6}
\nonumber
\end{eqnarray}

The overall sign of the new YCC can be reversed and the new overall sign can be
absorbed into $\delta$ and $\delta'$.
This can be done partially in the group of such constants that accompanies
$\delta$ or in the group that accompanies $\delta'$ or in both.
Because of this, we have determined only the absolute values of $\delta$ and
$\delta'$.
In order to emphasize this fact we have inserted absolute value bars on
$\delta$ and $\delta'$ in Eq.~(\ref{ang}).

The relevance of the values of the a priori mixing angles lies in that a
crucial test of the whole a priori mixings in hadrons scheme is that these
angles must show a universality-like proprety.
This is essential for this scheme to serve as a serious framework to describe
the enhancement phenomenon observed in non-leptonic, weak radiative, and
rare-mode weak decays of hadrons.

\acknowledgments

The authors wish to acknowledge partial support by CONACyT (M\'exico).

\begin{table}
\caption{
Experimental data on NLDH~[4].
The corresponding experimental values of the transition amplitudes are quoted
in the last two columns.
Only their absolute values are given, their signs are found in
Table~\ref{tablaii}.
}~
\label{tablai}
\begin{tabular}
{
c
r@{$\rightarrow$}l
r@{.}l@{\,$\pm$\,}r@{.}l
r@{.}l@{\,$\pm$\,}r@{.}l
r@{.}l@{\,$\pm$\,}r@{.}l
r@{.}l@{\,$\pm$\,}r@{.}l
r@{.}l@{\,$\pm$\,}r@{.}l
}
$i$ &
\multicolumn{2}{c}{Decay} &
\multicolumn{4}{c}{$\Gamma_i$} &
\multicolumn{4}{c}{$\alpha_i$} &
\multicolumn{4}{c}{$\gamma_i$} &
\multicolumn{4}{c}{$|A_i|$} &
\multicolumn{4}{c}{$|B_i|$}
\\
&
\multicolumn{2}{c}{} &
\multicolumn{4}{c}{($10^8 {\rm \ sec}^{-1}$)} &
\multicolumn{4}{c}{} & 
\multicolumn{4}{c}{} &
\multicolumn{4}{c}{($10^{-7}$)} &
\multicolumn{4}{c}{($10^{-7}$)}
\\
\tableline
1 &
$\Lambda$ & $p\pi^-$ &
24 & 28 & 0 & 26 &
0 & 642 & 0 & 013 &
0 & 76 & 0 & 01 &
3 & 231 & 0 & 018 &
22 & 15 & 0 & 36
\\
2 &
$\Lambda$& $n\pi^0$ &
13 & 60 & 0 & 22 &
0 & 65 & 0 & 05 &
0 & 76 & 0 & 04 &
2 & 374 & 0 & 027 &
15 & 88 & 0 & 98
\\
3 &
$\Sigma^-$ & $n\pi^-$ &
67 & 51 & 0 & 50 &
$-$0 & 068 & 0 & 008 &
0 & 98 & 0 & 05 &
4 & 269 & 0 & 016 &
1 & 43 & 0 & 17
\\
4 &
$\Sigma^+$ & $n\pi^+$ &
60 & 45 & 0 & 48 &
0 & 068 & 0 & 013 &
$-$0 & 97 & 0 & 08 &
0 & 140 & 0 & 027 &
42 & 17 & 0 & 16
\\
5 &
$\Sigma^+$ & $p\pi^0$ &
64 & 54 & 0 & 50 &
$-$0 & \multicolumn{3}{l}{$98{\ \ }^{+\ 0.017}_{-\ 0.015}$} &
0 & \multicolumn{3}{l}{$16{\ }^{+\ 0.10}_{-\ 0.09}$} &
3 & \multicolumn{3}{l}{$240{\ }^{+\ 0.082}_{-\ 0.100}$} &
26 & \multicolumn{3}{l}{$95{\ }^{+\ 1.16}_{-\ 1.02}$}
\\
6 &
$\Xi^-$ & $\Lambda\pi^-$ &
60 & 94 & 0 & 56 &
$-$0 & 456 & 0 & 014 &
0 & 89 & 0 & 01 &
4 & 497 & 0 & 023 &
17 & 47 & 0 & 48
\\
7 &
$\Xi^0$ & $\Lambda\pi^0$ &
34 & 32 & 1 & 07 &
$-$0 & 411 & 0 & 022 &
0 & 85 & 0 & 07 &
3 & 431 & 0 & 055 &
12 & 32 & 0 & 70
\\
\end{tabular}
\end{table}

\begin{table}
\caption{
Possible choices of the signs of the $A$'s and $B$'s.
Each choice is refered to by a subindex in the first column.
The negative sign of this subindex means that all signs are reversed 
w.\ r.\ t.\ the positive sign of the subindex.
}~
\label{tablaii}
\begin{tabular}
{
ccccccccccccccc
}
Choice of signs &
$B_1$ &
$B_2$ &
$B_3$ &
$B_4$ &
$B_5$ &
$B_6$ &
$B_7$ &
$A_1$ &
$A_2$ &
$A_3$ &
$A_4$ &
$A_5$ &
$A_6$ &
$A_7$ 
\\
\tableline
$SB_1$ $SA_1$ &
$+$ &
$-$ &
$-$ &
$+$ &
$+$ &
$-$ &
$-$ &
$+$ &
$-$ &
$+$ &
$+$ &
$-$ &
$+$ &
$+$
\\
$SB_2$ $SA_2$ &
$+$ &
$-$ &
$+$ &
$-$ &
$-$ &
$+$ &
$+$ &
$+$ &
$-$ &
$-$ &
$-$ &
$+$ &
$-$ &
$-$
\\
$SB_3$ $SA_3$ &
$+$ &
$-$ &
$+$ &
$-$ &
$-$ &
$-$ &
$-$ &
$+$ &
$-$ &
$-$ &
$-$ &
$+$ &
$+$ &
$+$
\\
$SB_4$ $SA_4$ &
$+$ &
$-$ &
$-$ &
$+$ &
$+$ &
$+$ &
$+$ &
$+$ &
$-$ &
$+$ &
$+$ &
$-$ &
$-$ &
$-$
\\
$SB_{-1}$ $SA_{-1}$ &
$-$ &
$+$ &
$+$ &
$-$ &
$-$ &
$+$ &
$+$ &
$-$ &
$+$ &
$-$ &
$-$ &
$+$ &
$-$ &
$-$
\\
$SB_{-2}$ $SA_{-2}$ &
$-$ &
$+$ &
$-$ &
$+$ &
$+$ &
$-$ &
$-$ &
$-$ &
$+$ &
$+$ &
$+$ &
$-$ &
$+$ &
$+$
\\
$SB_{-3}$ $SA_{-3}$ &
$-$ &
$+$ &
$-$ &
$+$ &
$+$ &
$+$ &
$+$ &
$-$ &
$+$ &
$+$ &
$+$ &
$-$ &
$-$ &
$-$
\\
$SB_{-4}$ $SA_{-4}$ &
$-$ &
$+$ &
$+$ &
$-$ &
$-$ &
$-$ &
$-$ &
$-$ &
$+$ &
$-$ &
$-$ &
$+$ &
$+$ &
$+$
\\
\end{tabular}
\end{table}

\narrowtext

\begin{table}
\caption{
Experimental values of the YCC we shall need later on.
The signs shown correspond to the signs expected from strong-flavor $SU_3$.
}~
\label{tablaiii}
\begin{tabular}
{
c
r@{.}l@{\,$\pm$\,}r@{.}l
}
$g_{{}_{B,B'M}}$ &
\multicolumn{4}{c}{Experimental value}
\\
\tableline
$g_{{}_{p,p\pi^0}}$ &
1 & 0000 & 0 & 0063
\\
$g_{{}_{\Sigma^+,\Lambda\pi^+}}(=g_{{}_{\Lambda,\Sigma^+\pi^-}})$ &
$-$0 & 897 & 0 & 074
\\
$g_{{}_{\Sigma^+,\Sigma^+\pi^0}}$ &
0 & 936 & 0 & 075
\\
$g_{{}_{p,\Sigma^0K^+}}(=g_{{}_{\Sigma^0,pK^-}})$ &
0 & 251 & 0 & 056
\\
$g_{{}_{p,\Lambda K^+}}(=g_{{}_{\Lambda,pK^-}})$ &
0 & 987 & 0 & 092
\\
$g_{{}_{\Xi^0,\Xi^0\pi^0}}$ &
$-$0 & 270 & 0 & 081
\\
$g_{{}_{\Xi^-,\Lambda K^-}}$ &
$-$0 & 266 & 0 & 080
\\
\end{tabular}
\end{table}

\widetext

\begin{table}
\caption{
Predictions obtained with Eqs.~(\ref{bes}) for the $B$'s.
The values of the YCC are also displayed.
The last line displays the corresponding value of $\sigma$.
The subindeces in the headings of the columns indicate the choice of the signs
of the experimental $B$'s according to Table~\ref{tablaii}.
}~
\label{tablaiv}
\begin{tabular}
{
cdddd
}
&
$SB_2$ &
$SB_3$ &
$SB_{-1}$ &
$SB_{-4}$ 
\\
\tableline
$g_{{}_{p,p\pi^0}}$ &
1.00 &
1.00 &
1.00 &
1.00
\\
$g_{{}_{\Sigma^+,\Lambda\pi^+}}$ &
$-$0.86 &
$-$0.82 &
$-$0.78 &
$-$0.77
\\
$g_{{}_{\Sigma^+,\Sigma^+\pi^0}}$ &
0.94 &
0.92 &
0.88 &
0.88
\\
$g_{{}_{p,\Sigma^0K^+}}$ &
0.28 &
0.26 &
0.24 &
0.23
\\
$g_{{}_{p,\Lambda K^+}}$ &
1.03 &
1.04 &
0.87 &
0.89
\\
$g_{{}_{\Xi^0,\Xi^0\pi^0}}$ &
$-$0.27 &
$-$0.35 &
$-$0.26 &
$-$0.31
\\
$g_{{}_{\Xi^-,\Lambda K^-}}$ &
$-$0.27 &
$-$0.31 &
$-$0.26 &
$-$0.28
\\
$B_1\ (10^{-7})$ &
22.17 &
22.17 &
$-$22.17 &
$-$22.17
\\
$B_2\ (10^{-7})$ &
$-$15.68 &
$-$15.68 &
15.67 &
15.68
\\
$B_3\ (10^{-7})$ &
1.37 &
1.37 &
1.37 &
1.37
\\
$B_4\ (10^{-7})$ &
$-$42.12 &
$-$42.12 &
$-$42.12 &
$-$42.12
\\
$B_5\ (10^{-7})$ &
$-$30.75 &
$-$30.76 &
$-$30.76 &
$-$30.75
\\
$B_6\ (10^{-7})$ &
17.46 &
$-$17.45 &
17.46 &
$-$17.46
\\
$B_7\ (10^{-7})$ &
12.35 &
$-$12.34 &
12.35 &
$-$12.35
\\
$\sigma\ (10^{-6})$ &
\multicolumn{1}{c}{$3.9\pm 1.1$} &
\multicolumn{1}{c}{$5.0\pm 1.6$} &
\multicolumn{1}{c}{$7.3\pm 2.7$} &
\multicolumn{1}{c}{$8.1\pm 3.1$}
\\
\end{tabular}
\end{table}

\begin{table}
\caption{
Predictions for the $A$'s and values of the YCC that accompany them.
The values of the angles $\delta$ and $\delta'$ are given in the last two
lines.
The headings indicate the signs of the $A$'s, as in Table~\ref{tablaii}, and
the signs of the new YCC.
The order of the successions of 0's and 1's follow the order of
the YCC in the lines of this table, 0 means same sign and 1 opposite sign
with respect to the corresponding lines of Table~\ref{tablaiii}.
Notice that there are eight new YCC while there were seven ordinary ones.
The eighth is $g^{{}^{p,sp}}_{{}_{\Sigma^+,\Lambda\pi^+}}$, its relative sign
with respect to $g_{{}_{\Sigma^+,\Lambda\pi^+}}$ of Table~\ref{tablaiii} is
given by the last (eighth) digit in the successions in the headings.
}~
\label{tablav}
\begin{tabular}
{
cdddd
}
&
$SA_{-4}$ &
$SA_{-1}$ &
$SA_{-4}$ &
$SA_4$ 
\\
&
10010011 &
10010100 &
10011011 &
01100100
\\
\tableline
$g^{{}^{p,sp}}_{{}_{p,p\pi^0}}$ &
$-$1.00 &
$-$1.00 &
$-$1.00 &
1.00
\\
$g^{{}^{s,pp}}_{{}_{\Sigma^+,\Lambda\pi^+}}$ &
$-$0.82 &
$-$0.82 &
$-$1.20 &
1.20
\\
$g^{{}^{s,pp}}_{{}_{\Sigma^+,\Sigma^+\pi^0}}$ &
1.03 &
1.04 &
0.49 &
$-$0.49
\\
$g^{{}^{s,ss}}_{{}_{p,\Sigma^0K^+}}$ &
$-$0.22 &
$-$0.22 &
$-$0.42 &
0.42
\\
$g^{{}^{s,ss}}_{{}_{p,\Lambda K^+}}$ &
0.83 &
0.84 &
$-$0.35 &
0.36
\\
$g^{{}^{s,pp}}_{{}_{\Xi^0,\Xi^0\pi^0}}$ &
$-$0.35 &
0.35 &
$-$0.041 &
0.041
\\
$g^{{}^{s,ss}}_{{}_{\Xi^-,\Lambda K^-}}$ &
0.22 &
$-$0.22 &
0.39 &
$-$0.39
\\
$g^{{}^{p,sp}}_{{}_{\Sigma^+,\Lambda\pi^+}}$ &
0.82 &
$-$0.82 &
1.20&
$-$1.20
\\
$A_1\ (10^{-7})$ &
$-$3.25 &
$-$3.25 &
$-$3.26&
3.26
\\
$A_2\ (10^{-7})$ &
2.30 &
2.30 &
2.31&
$-$2.31
\\
$A_3\ (10^{-7})$ &
$-$4.28 &
$-$4.27 &
$-$4.27&
4.27
\\
$A_4\ (10^{-7})$ &
$-$0.123 &
$-$0.123 &
$-$0.145&
0.145
\\
$A_5\ (10^{-7})$ &
2.94 &
2.94 &
2.92 &
$-$2.92
\\
$A_6\ (10^{-7})$ &
4.53 &
$-$4.53 &
4.52 &
$-$4.52
\\
$A_7\ (10^{-7})$ &
3.20 &
$-$3.20 &
3.20 &
$-$3.20
\\
$\delta\ (10^{-6})$ &
\multicolumn{1}{c}{$0.20\pm 0.03$} &
\multicolumn{1}{c}{$0.20\pm 0.03$} &
\multicolumn{1}{c}{$0.066\pm 0.004$} &
\multicolumn{1}{c}{$0.066\pm 0.004$}
\\
$\delta'\ (10^{-6})$ &
\multicolumn{1}{c}{$0.23\pm 0.03$} &
\multicolumn{1}{c}{$0.24\pm 0.03$} &
\multicolumn{1}{c}{$0.082\pm 0.004$} &
\multicolumn{1}{c}{$0.082\pm 0.004$}
\\
\end{tabular}
\end{table}

\narrowtext

\begin{table}
\caption{
Predictions of Eqs.~(\ref{aes}) and (\ref{bes}) for the experimental
observables.
The headings in parentheses indicate the signs that result for
the predictions for $A$'s and $B$'s, as in Table~\ref{tablaii}.
The $\Gamma_i$ are in $10^8$~sec$^{-1}$.
}~
\label{tablavi}
\begin{tabular}
{
cddd
}
&
$(SB_{-4}$ $SA_{-4})$ &
$(SB_{-1}$ $SA_{-1})$ &
$(SB_{-4}$ $SA_{-4})$
\\
\tableline
$\Gamma_1$ &
24.76&
24.77&
24.82\\
$\Gamma_2$ &
12.92&
12.92&
12.95\\
$\Gamma_3$ &
67.35&
67.36&
67.35\\
$\Gamma_4$ &
60.22&
60.22&
60.25\\
$\Gamma_5$ &
64.94&
64.94&
64.87\\
$\Gamma_6$ &
61.52&
61.53&
61.45\\
$\Gamma_7$ &
29.97&
29.97&
29.94\\
$\alpha_1$ &
0.646&
0.646&
0.646\\
$\alpha_2$ &
0.66&
0.66&
0.66\\
$\alpha_3$ &
$-$0.064&
$-$0.064&
$-$0.065\\
$\alpha_4$ &
0.074&
0.074&
0.077\\
$\alpha_5$ &
$-$1.00&
$-$1.00&
$-$1.00\\
$\alpha_6$ &
$-$0.449&
$-$0.449&
$-$0.449\\
$\alpha_7$ &
$-$0.438&
$-$0.438&
$-$0.438\\
$\gamma_1$ &
0.76&
0.76&
0.76\\
$\gamma_2$ &
0.75&
0.75&
0.75\\
$\gamma_3$ &
1.00&
1.00&
1.00\\
$\gamma_4$ &
$-$1.00&
$-$1.00&
$-$1.00\\
$\gamma_5$ &
$-$0.05&
$-$0.05&
$-$0.05\\
$\gamma_6$ &
0.89&
0.89&
0.89\\
$\gamma_7$ &
0.90&
0.90&
0.90\\
\end{tabular}
\end{table}

\mediumtext

\begin{table}
\caption{
Values of the YCC and of the a priori mixing angles that correspond to the
predictions of Table~\ref{tablavi}.
The headings are explained in the captions of Tables~\ref{tablavi} and
\ref{tablav}.
}~
\label{tablavii}
\begin{tabular}
{
cddd
}
&
$(SB_{-4}$ $SA_{-4})$ &
$(SB_{-1}$ $SA_{-1})$ &
$(SB_{-4}$ $SA_{-4})$
\\
&
10010011 &
10010100 &
10011011
\\
\tableline
$g_{{}_{p,p\pi^0}}$ &
1.00&
1.00&
0.99\\
$g_{{}_{\Sigma^+,\Lambda\pi^+}}$ &
$-$0.81&
$-$0.83&
$-$1.11\\
$g_{{}_{\Sigma^+,\Sigma^+\pi^0}}$ &
0.99&
0.99&
0.62\\
$g_{{}_{p,\Sigma^0K^+}}$ &
0.22&
0.23&
0.66\\
$g_{{}_{p,\Lambda K^+}}$ &
0.77&
0.75&
0.41\\
$g_{{}_{\Xi^0,\Xi^0\pi^0}}$ &
$-$0.36&
$-$0.29&
$-$0.19\\
$g_{{}_{\Xi^-,\Lambda K^-}}$ &
$-$0.30&
$-$0.20&
$-$0.93\\
$\sigma\ (10^{-6})$ &
\multicolumn{1}{c}{$4.9\pm 1.5$}&
\multicolumn{1}{c}{$4.5\pm 1.3$}&
\multicolumn{1}{c}{$3.0\pm 0.9$}\\
$\delta\ (10^{-6})$ &
\multicolumn{1}{c}{$0.23\pm 0.07$} &
\multicolumn{1}{c}{$0.21\pm 0.07$}&
\multicolumn{1}{c}{$0.061\pm 0.001$}\\
$\delta'\ (10^{-6})$ &
\multicolumn{1}{c}{$0.26\pm 0.07$} &
\multicolumn{1}{c}{$0.25\pm 0.06$}&
\multicolumn{1}{c}{$0.076\pm 0.005$}\\
\end{tabular}
\end{table}

\narrowtext

\begin{table}
\caption{
Effect of $SU(2)$ breaking in the predictions for the observables.
The $\Gamma_i$ are in $10^8$~sec$^{-1}$.
This table should be compared with Table~\ref{tablavi}.
}~
\label{tablaviii}
\begin{tabular}
{
cddd
}
&
$(SB_{-4}$ $SA_{-4})$ &
$(SB_{-1}$ $SA_{-1})$ &
$(SB_{-4}$ $SA_{-4})$
\\
\tableline
$\Gamma_1$ &
24.30&
24.30&
24.32\\
$\Gamma_2$ &
13.57&
13.57&
13.60\\
$\Gamma_3$ &
67.53&
67.53&
67.56\\
$\Gamma_4$ &
60.45&
60.46&
60.45\\
$\Gamma_5$ &
64.52&
64.52&
64.48\\
$\Gamma_6$ &
60.95&
60.95&
60.94\\
$\Gamma_7$ &
34.27&
34.28&
34.00\\
$\alpha_1$ &
0.645&
0.645&
0.646\\
$\alpha_2$ &
0.67&
0.67&
0.66\\
$\alpha_3$ &
$-$0.068&
$-$0.068&
$-$0.068\\
$\alpha_4$ &
0.067&
0.067&
0.068\\
$\alpha_5$ &
$-$0.98&
$-$0.98&
$-$0.99\\
$\alpha_6$ &
$-$0.456&
$-$0.456&
$-$0.456\\
$\alpha_7$ &
$-$0.414&
$-$0.414&
$-$0.412\\
$\gamma_1$ &
0.76&
0.76&
0.76\\
$\gamma_2$ &
0.74&
0.74&
0.76\\
$\gamma_3$ &
1.00&
1.00&
1.00\\
$\gamma_4$ &
$-$1.00&
$-$1.00&
$-$1.00\\
$\gamma_5$ &
0.18&
0.18&
0.11\\
$\gamma_6$ &
0.89&
0.89&
0.89\\
$\gamma_7$ &
0.91&
0.91&
0.91\\
\end{tabular}
\end{table}

\widetext

\begin{table}
\caption{
$SU(2)$ breaking corrections to the YCC and their effect upon the a priori
mixing angles.
We show the predictions for the experimentally reported YCC of
Table~\ref{tablaiii} and, in order to facilitate the visualization of the
changes of the additional YCC from the $SU(2)$ symmetry limit,
we show the ratios of these additional constants to their
symmetry limit, except for $g_{{}_{\Lambda,\Lambda\pi^0}}$ which is zero in
that limit and for it we display the constant itself.
In the symmetry limit the ratios are of course 1.
Since the breakings are small, the $SU(2)$ relative signs of the YCC are
preserved.
The signs of the new YCC are indicated by the successions of 0's and 1's in the
headings.
We use only 10 of these digits and not 23.
Then each of the first 7 digits (from left to right) gives the relative sign
between the group of the new YCC that belong the same isomultiplets and the
corresponding ordinary YCC of Table~\ref{tablaiii}.
The eighth digit gives the relative sign of the group
$g^{p,sp}_{{}_{\Sigma^+,\Lambda\pi^+}}$,
$g^{p,sp}_{{}_{\Sigma^0,\Lambda\pi^0}}$,
and
$g^{p,sp}_{{}_{\Sigma^-,\Lambda\pi^-}}$
w.\ r.\ t.\ the ordinary
$g_{{}_{\Sigma^+,\Lambda\pi^+}}$.
The ninth and tenth digits give the relative signs between
$g^{s,pp}_{{}_{\Lambda,\Lambda\pi^0}}$
and
$g^{p,sp}_{{}_{\Lambda,\Lambda\pi^0}}$
w.\ r.\ t.\ the ordinary
$g_{{}_{\Lambda,\Lambda\pi^0}}$,
respectively.
The results in this table should be compared with those in
Table~\ref{tablavii}.
}~
\label{tablaix}
\begin{tabular}
{
lddd
}
&
$(SB_{-4}$ $SA_{-4})$ &
$(SB_{-1}$ $SA_{-1})$ &
$(SB_{-4}$ $SA_{-4})$
\\
&
1001001100 &
1001010001 &
1001101101
\\
\tableline
$g_{{}_{p,p\pi^0}}$ &
1.00 &
1.00 &
1.00 \\
$g_{{}_{n,n\pi^0}}/(-g_{{}_{p,p\pi^0}})$ &
1.02 &
1.03 &
0.96 \\
$g_{{}_{p,n\pi^+}}/\sqrt{2}g_{{}_{p,p\pi^0}}$ &
0.98 &
0.98 &
1.00 \\
$g_{{}_{n,p\pi^-}}/\sqrt{2}g_{{}_{p,p\pi^0}}$ &
1.00 &
1.01 &
0.94 \\
$g_{{}_{\Sigma^+,\Lambda\pi^+}}$ &
$-$0.82 &
$-$0.86 &
$-$1.07 \\
$g_{{}_{\Sigma^0,\Lambda\pi^0}}/g_{{}_{\Sigma^+,\Lambda\pi^+}}$ &
1.01 &
1.01 &
0.98 \\
$g_{{}_{\Sigma^-,\Lambda\pi^-}}/g_{{}_{\Sigma^+,\Lambda\pi^+}}$ &
0.98 &
0.98 &
0.95 \\
$g_{{}_{\Lambda,\Sigma^+\pi^-}}/g_{{}_{\Sigma^+,\Lambda\pi^+}}$ &
0.98 &
0.98 &
1.03 \\
$g_{{}_{\Lambda,\Sigma^0\pi^0}}/g_{{}_{\Sigma^+,\Lambda\pi^+}}$ &
1.01 &
1.01 &
1.03 \\
$g_{{}_{\Sigma^+,\Sigma^+\pi^0}}$ &
0.95 &
0.94 &
0.63 \\
$g_{{}_{\Sigma^+,\Sigma^0\pi^+}}/(-g_{{}_{\Sigma^+,\Sigma^+\pi^0}})$ &
1.00 &
1.01 &
1.00 \\
$g_{{}_{\Sigma^-,\Sigma^0\pi^-}}/g_{{}_{\Sigma^+,\Sigma^+\pi^0}}$ &
1.01 &
1.01 &
1.02 \\
$g_{{}_{p,\Sigma^0K^+}}$ &
0.23 &
0.28 &
0.57 \\
$g_{{}_{\Sigma^-,nK^-}}/\sqrt{2}g_{{}_{p,\Sigma^0K^+}}$ &
0.98 &
0.98 &
0.99 \\
$g_{{}_{\Sigma^+,p\bar{K}^0}}/\sqrt{2}g_{{}_{p,\Sigma^0K^+}}$ &
1.02 &
1.02 &
1.05 \\
$g_{{}_{p,\Lambda K^+}}$ &
0.80 &
0.74 &
0.36 \\
$g_{{}_{\Lambda,pK^-}}/g_{{}_{p,\Lambda K^+}}$ &
0.98 &
0.98 &
0.97 \\
$g_{{}_{\Lambda,n\bar{K}^0}}/g_{{}_{p,\Lambda K^+}}$ &
1.01 &
1.01 &
1.01 \\
$g_{{}_{\Xi^0,\Xi^0\pi^0}}$ &
$-$0.35 &
$-$0.27 &
$-$0.20\\
$g_{{}_{\Xi^-,\Xi^0\pi^-}}/\sqrt{2}g_{{}_{\Xi^0,\Xi^0\pi^0}}$ &
1.01 &
1.00 &
0.99 \\
$g_{{}_{\Xi^-,\Lambda K^-}}$ &
$-$0.32 &
$-$0.23 &
$-$0.80 \\
$g_{{}_{\Xi^0,\Lambda\bar{K}^0}}/(-g_{{}_{\Xi^-,\Lambda K^-}})$ &
1.01 &
1.00 &
1.07 \\
$g_{{}_{\Lambda,\Lambda\pi^0}}$ &
0.012 &
0.014 &
$-$0.014 \\
$\sigma\ (10^{-6})$ &
\multicolumn{1}{c}{$4.8\pm 1.5$} &
\multicolumn{1}{c}{$4.9\pm 2.0$} &
\multicolumn{1}{c}{$4.9\pm 2.5$} \\
$\delta\ (10^{-6})$ &
\multicolumn{1}{c}{$0.22\pm 0.07$} &
\multicolumn{1}{c}{$0.22\pm 0.09$} &
\multicolumn{1}{c}{$0.075\pm 0.007$} \\
$\delta'\ (10^{-6})$ &
\multicolumn{1}{c}{$0.26\pm 0.08$} &
\multicolumn{1}{c}{$0.26\pm 0.09$} &
\multicolumn{1}{c}{$0.089\pm 0.010$} \\
\end{tabular}
\end{table}


\begin{references}

\bibitem{LE-6129}
A.~Garc\'{\i}a, R.~Huerta, and G.~S\'anchez-Col\'on,
``{\it A priori} mixed hadrons, weak radiative and non-leptonic decays of hyperons".
e-print archive hep-ph/9804341,
to be published in J. of Phys. {\bf G}.

\bibitem{marshak}
Original references for the $|\Delta I|=1/2$ rule can be found in
R.~E.~Marshak, Riazuddin, and C.~P.~Ryan, {\it The Theory of Weak Interactions
in Particle Physics} (Wiley-Interscience, John Wiley and Sons, Inc., 1969).

\bibitem{dumbrajs}
O.~Dumbrajs {\it et al.}, Nucl.\ Phys.\ {\bf B 216}, 277 (1983).

\bibitem{pdg}
For a review of particle properties, see L.~Montanet {\it et al.} Phys.\ Rev.\ {\bf D 50}, 1173 (1994).

\end{references}
\end{document}